\documentclass{svproc}
\usepackage{url}
\usepackage{graphicx}
\usepackage{amsmath}
\usepackage{amssymb}
\usepackage{upgreek}
\usepackage{float}
\usepackage{multicol}
\usepackage{footmisc}

\def\bfx{{ \bf x  }}
\def\bfX{{\bf X}}
\def\defeq{{\stackrel{\Delta}{=}}}
\def\calX{{\mathcal{X}}}

\begin{document}
\mainmatter              
\title{Local Conditioning:\\ \smallskip {\normalsize Exact Message Passing\\ \vspace{-2mm} for Cyclic Undirected Distributed Networks}}
\titlerunning{Local Conditioning in Undirected Networks}  
%

\author{Matthew G. Reyes}
\authorrunning{MGR} 
%
\tocauthor{Matthew G. Reyes}
\institute{Independent Researcher and Consultant\\
Ann Arbor, MI 48105, USA\\
\email{matthewgreyes@yahoo.com}\\ 
\texttt{matthewgreyes.com}
}


\maketitle              

\begin{abstract}
This paper addresses practical implementation of {\em summing out}, {\em expanding}, and {\em reordering} of messages in Local Conditioning (LC) for undirected networks. In particular, incoming messages conditioned on potentially different subsets of the receiving node's relevant set must be {\em expanded} to be conditioned on this relevant set, then {\em reordered} so that corresponding columns of the conditioned matrices can be fused through element-wise multiplication. An outgoing message is then reduced by {\em summing out} loop cutset nodes that are {\em upstream} of the outgoing edge. The emphasis on implementation is the primary contribution over the theoretical justification of LC given in Fay et al. Nevertheless, the complexity of Local Conditioning in grid networks is still no better than that of Clustering.
\end{abstract}
\section{Introduction}\label{sec:introduction}

Local Conditioning (LC) has recently been proposed as a method for performing exact truly distributed inference in cyclic undirected networks \cite{reyes2017}. Originally introduced in the context of directed networks \cite{diez}, \cite{fay}, LC ceased to gain attention in favor of Clustering methods such as the well-known Junction Tree algorithm \cite{lauritzen} and generalizations thereof \cite{yedidia}. This was likely due both to the relative complexity of implementing Local Conditioning, and a surge of interest in approximate variations of Belief Propagation for cyclic networks \cite{frey}, \cite{murphy}. Moreover, the principle advantage of Local Conditioning over Clustering methods was not as pressing as it is today. That is to say, Local Conditioning is {\em truly distributed} in the sense of involving messages between individual nodes of the original network. Truly distributed inference is required by physically distributed systems such as autonomous vehicles, industrial warehouse sensors, or networks of delivery drones, for which clustering of nodes is not feasible. Indeed, it is precisely the current preponderance of physically distributed computing that motivates a renewed interest in Local Conditioning. This paper develops Local Conditioning in the case of undirected cyclic networks. The simpler topological structure of undirected networks and the economy of Gibbs distributions permits a more streamlined presentation than in the case of Bayesian networks \cite{fay}. Moreover, the discussion is more detailed than earlier work \cite{reyes2017} and lays particular emphasis on the implementation of Local Conditioning.

{\em Belief Propagation} (BP) on an acyclic network \cite{pearl} consists of two main ideas, {\em fusion} and {\em propagation}. Each of these can be interpreted as an appropriately defined multiplication of matrices: fusion as an element-wise product of two or more incoming message vectors; propagation as standard matrix multiplication of a fused vector with a matrix associated with an outgoing edge. It is important to distinguish between two senses in which BP can be said to be a {\em distributed} algorithm. The first is as formalized in the {\em Generalized Distributive Law} \cite{aji}, which presents Belief Propagation as a means of solving an otherwise intractable summation\footnote{The reader should consult \cite{aji} for the algebraic equivalence of other operations.} by decomposing it into a sequence of much smaller computations. In this sense, BP is a {\em distributed algorithm} even if it is implemented on a centralized computer. This is significant as by far the most common means of adapting BP for performing exact inference on cyclic networks is to form an acyclic network by clustering nodes \cite{lauritzen}. This clustering is possible precisely when BP is implemented on a centralized computer. 

{\em Conditioning} \cite{pearl0}, \cite{suermondt}, is an adaption of BP for performing exact inference in a cyclic network by effectively `opening up' the network. Conditioning was introduced around the same time as Clustering methods. It consists of the same two ideas of fusion and propagation, only instead of vector messages there are matrix messages, each column of which corresponds to, i.e., is conditioned on, a different configuration of a set of so-called {\em loop cutset} nodes. Conditioning on the loop cutset nodes allows standard acyclic Belief Propagation to be implemented on a cyclic network, by having each loop cutset node interact with its neighbors {\em as if} there are multiple copies of it, all constrained to have the same value. The multiple copies of a loop cutset node effectively break the cycles going through the node. The constraint that individual copies have the same value results in the increase in the number of message vectors that need to be passed, so that messages corresponding to different common assignments can be aligned.
 
{\em Local Conditioning} \cite{diez}, \cite{fay}, \cite{reyes2017}, reduces the number of columns that need to be included in the message matrices by noting that a message matrix passed over a given edge need only be conditioned on a particular loop cutset node if there is at least one copy of the loop cutset on either side of the edge. In addition to the standard operations of fusion and propagation, there are three additional operations necessary for implementation: {\em summing out}, {\em expanding}, and {\em reordering}. Summing out loop cutset nodes that are {\em upstream} of a message decreases the number of columns in the message matrix and is part of the propagation step. As a result, incoming messages to a given node will be conditioned on potentially different subsets of the loop cutset nodes. Therefore, incoming message matrices need to be expanded to a common subset. The expanded message matrices then need to be reordered so that corresponding columns of the incoming messages in turn correspond to the same configuration on the common subset.

This paper addresses practical implementation of {\em summing out}, {\em expanding}, and {\em reordering} for Local Conditioning. In addition, it discusses the {\em associated tree} and {\em stability scaling}, or normalization, important for the practical implementation of general Conditioning. Stability scaling is used in practice with standard Belief Propagation as a means of preventing numerical overflow or underflow in large networks. Scaling the messages of BP does not affect probabilities computed therefrom. However, in Conditioning, columns of the matrix messages are conditioned on different configurations of loop cutset nodes. In order to combine columns of scaled messages in a way that still yields correct probabilities for the cyclic network, messages corresponding to different loop cutset configurations must be scaled with a common factor rather than independently. The present paper does not address finding such a common factor for scaling.


This paper is organized as follows. Section \ref{sec:goal} provides an overall goal statement. Sections \ref{sec:gibbs} and \ref{sec:bp} provide background on Gibbs distributions and Belief Propagation, respectively. Section \ref{sec:cycles} provides a brief account of techniques to address cycles in networks. Section \ref{sec:conditioning} provides a discussion of node-specific message passing for Conditioning. Section \ref{sec:local} presents the development of Local Conditioning for cyclic undirected networks. Section \ref{sec:conclusion} concludes with future direction.

\vspace{2mm}
\section{Goal Statement: Single-Node Inference on Networks}\label{sec:goal}

The {\em inference} addressed in this paper is the computation of probabilities at individual nodes in the network. Such probabilities will be used by nodes for making individual decisions.  In general such probabilities will be inferred conditioned upon observed data. However, such conditioning can be incorporated into the framework of an unconditioned model, which is presented here for simplicity.  

A network $G = (V,E)$ consists of a set of {\em nodes} $V$ and a set of {\em edges} $E$ consisting of pairs of elements of $V$. For edge $\{i,j\} \in E$, nodes $i$ and $j$ are said to be {\em neighbors}. The graph $G \setminus \{i,j\}$ is obtained by removing the edge between $i$ and $j$. For node $i$, $\partial i$ denotes the set of neighbors of $i$. Given a neighbor $j\in\partial i$, the set $k\in\partial j\setminus i$ denotes the set of neighbors of $j$ not including $i$. For each $i\in V$, associate a random variable $X_i$ assuming values in a common {\em alphabet} $\calX$. Let $x_i$ denote a specific value that $X_i$ assumes. An assignment $\bfx = (x_1,\ldots,x_{|V|})$ to all nodes in the network is referred to as a {\em configuration}, and $\bfX$ denotes the {\em random field} of possible configurations taking values $\calX^{|V|}$. For a subset $L \subset V$, $\bfX_L$ denotes the random field on $L$, and $\bfx_L$ a configuration on $L$. A {\em path} is a sequence of nodes $k, \ldots, i$ such that any two successive nodes in the sequence are neighbors, and each such pair of neighbors occurs only once in the sequence. A {\em cycle} is a path that begins and ends at the same node. The goal of this paper is to compute, for each node $i\in V$, the vector of probabilities $p_i(x_i)$ for each value $x_i$ that node $i$ can assume. In particular, we want to find an algorithm for computing these probabilities when the network $G$ has cycles and corresponds to a physically distributed system requiring truly distributed inference.

\begin{figure}[t]
	\begin{center}
		\hbox{
			\hspace{1.38in}
			\includegraphics[scale=.27]{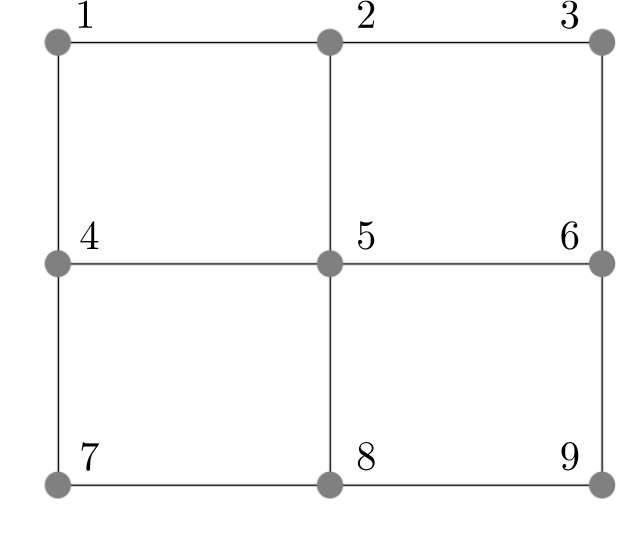}
		}
	\end{center}
	\vspace{-8mm}
	\caption{$3\times 3$ grid network.}
	\label{fig:grid_graph_1}
\end{figure}

As an example, consider a grid graph as shown in Figure \ref{fig:grid_graph_1}. The grid topology is a good first-order approximation to a number of distributed networks of practical interest. For example, autonomous vehicles will communicate with those that are close by as well as with base stations positioned along roadsides and medians; in addition, industrial sensors and robots will likely be positioned along warehouse rows and columns. Moreover, the simple geometric structure of grid graphs permits an exploration and articulation of basic principles that, once grasped, can be abstracted to more general topologies.

While the present discussion is couched in terms of probabilities, the contribution of this paper is ultimately in the graph-theoretic manipulation of a network to facilitate communication. As such, the ideas of this paper are applicable to settings in which information other than probabilities are desired.

\vspace{2mm}
\section{Gibbs Distributions}\label{sec:gibbs}

Probabilities will be computed with respect to {\em Gibbs} distributions. {\em Any} multivariate probability distribution that assigns non-zero probability to every configuration of the variables can be parametrized as a Gibbs distribution \cite{grimmet}. Gibbs distributions are {\em data-driven} in the sense of predicting observed data without making any additional assumptions beyond positivity \cite{cover}. Since one should not exclude the possibility of a configuration simply because the configuration has not yet been observed \cite{taleb}, the positivity assumption is a natural one to include in a model, and as such, Gibbs distributions are an extremely general class of models. Furthermore, their conditional independence structure admits a graphical interpretation that maps directly onto problems involving variables whose interdependence arises from communication or physical proximity.

For neighbors $i$ and $j$ there is an {\em edge potential}, $\Uppsi_{ij} = [\Uppsi_{ij}(x_i,x_j)]$, that maps each configuration $(x_i,x_j)$ to a positive number. Likewise, for each node $i$ there is a {\em self potential} $\Upphi_i = [\Upphi_i(x_i)]$ that assigns each value of $x_i$ to a positive number. This paper assumes a finite alphabet $\calX$, in which case the edge potential $\Uppsi_{ij}$ can be thought of as a matrix, the self potential $\Upphi_i$ a vector. To make things concrete, consider an {\em Ising} model, in which $\calX = \{-1,1\}$ and
\begin{align*}
    \Upphi_i & = \left[\begin{array}{c}
					e^{\alpha_i}\\
					e^{-\alpha_i}\\                                      
					\end{array}\right] ~~~~~~~~~~~~~~~ , & \Uppsi_{ij} & = \left[\begin{array}{cc}
										      e^{\theta_{ij}} & e^{-\theta_{ij}}\\
										      e^{-\theta_{ij}} & e^{\theta_{ij}}\\                                      
										      \end{array}\right] ,
\end{align*}
for parameters $\alpha_i$, $\theta_{ij}$.

The {\em belief} $Z_i$ for node $i$ is a vector with components $[Z_i(x_i)]$ defined as
\begin{eqnarray}
 Z_i(x_i) & = & \Upphi_i(x_i) \sum\limits_{\bfx_{V\setminus i}} \prod\limits_{j,k}\Uppsi_{jk}(x_j,x_k) \prod\limits_{j}\Upphi_j (x_j) . \nonumber
\end{eqnarray}

\vspace{0mm}
\section{Belief Propagation for Acyclic Networks}\label{sec:bp}

Let $\{i,j\}$ be an edge such that removing the edge between $i$ and $j$ disconnects the network. Belief Propagation (BP) is ultimately defined for {\em acyclic} networks, in which the removal of any edge disconnects the network. For such an edge, let $G_{i\setminus j}$ be the component of $G\setminus \{i,j\}$ that contains $i$. Furthermore, let $Z_i^{i\setminus j}$ be the belief for node $i$ with respect to the Gibbs distribution on $G_{i\setminus j}$ that inherits potentials from the original Gibbs distribution on $G$ in the natural way. It can be shown that
\begin{eqnarray}
  Z_i & ~ = ~ & Z_i^{i\setminus j} \Uppsi_{ji} Z_j^{j\setminus i} ~ . \label{eq:belief_decomp1}
\end{eqnarray}
We define the {\em message} from $j$ to $i$ as
\begin{eqnarray}
  m_{j\rightarrow i} & ~ \defeq ~ & \Uppsi_{ji} Z_j^{j\setminus i} ~ . \label{eq:message}
\end{eqnarray}
Note that this definition does not work if $\{i,j\}$ is not a cut edge. Node $k$ is said to be {\em upstream} of the message $m_{j\to i}$ if $k \in G_{j\setminus i}$. The message from $j$ to $i$ summarizes information about the potentials in $G_{j\setminus i}$ in that all potentials involving nodes in $G_{j\setminus i}$ have been {\em summed out}. Likewise, $k$ is said to be {\em downstream} of $m_{j\to i}$ if $k\in G_{i\setminus j}$. One can see that $Z_j^{j\setminus i}$, and therefore $m_{j\to i}$, is not a function of any downstream variables.  These simple observations are critical for the development of Local Conditioning in Section \ref{sec:local}.

It is helpful to think of BP as a {\em fusion} of message vectors incoming to a node and the self potential for that node, followed by {\em propagation} of the fused vector via multiplication with the corresponding edge potential matrix. Decomposing both $Z_i^{i\setminus j}$ and $Z_j^{j\setminus i}$ recursively according to (\ref{eq:belief_decomp1}) provides the respective formulas for beliefs and messages:
\begin{eqnarray}
  Z_i & = & \Upphi_i \prod\limits_{j\in\partial i}m_{j\rightarrow i}  \label{eq:belief_formula}
\end{eqnarray}
and
\begin{eqnarray}
    m_{j\rightarrow i} & = & \Uppsi_{ji} \Upphi_j \prod\limits_{k\in\partial j\setminus i}m_{k\rightarrow j} ~ . \label{eq:message_recursion}
\end{eqnarray}

Interest in the belief $Z_i$ is due to the fact that normalizing it gives the probabilities that node $i$ assumes value $-1$ or $1$. That is,
\begin{eqnarray}
  p_i(x) & = & \frac{ Z_i(x) }{ \sum\limits_{x\in \{-1,1\}} Z_i(x) } ~ . \nonumber
\end{eqnarray}
To prevent numerical overflow or underflow in large networks, the messages in (\ref{eq:message_recursion}) are scaled, or normalized. This does not affect computation of probabilities.



\vspace{2mm}
\section{Dealing with Cycles}\label{sec:cycles}

When a network has cycles, the message recursion of (\ref{eq:message_recursion}) will in general not result in correct computation of beliefs. Nevertheless, one can form an acyclic network by grouping the nodes of the original network into {\em cluster nodes} and creating an edge between two cluster nodes if each contains an endpoint of an edge in the original network \cite{lauritzen}. For example, one can cluster the grid network of Figure \ref{fig:grid_graph_1} into a chain network by creating clusters $c_1 = \{1,4,7\}$, $c_2 = \{2,5,8\}$, and $c_3 = \{3,6,9\}$, and edges $\{c_1,c_2\}$ and $\{c_2,c_3\}$. This approach is relatively straightforward if the algorithm is implemented on a centralized computer, where it is simply a matter of creating new variables with larger alphabets. However, if the nodes correspond to distributed units such as autonomous vehicles or industrial sensors, Clustering requires additional units with which clusters of individual units would need to be able to communicate. While such additional layers of infrastructure and communication may be feasible in some settings, it is important to consider {\em truly distributed} algorithms in which messages are passed between nodes of the original network.

Loopy Belief Propagation (LBP) \cite{murphy} is a truly distributed algorithm for performing {\em approximate} inference. In LBP nodes follow the message recursion of (\ref{eq:message_recursion}) {\em as if} they were part of an acyclic network. While LBP has been extensively studied, there is as yet little understanding of what exactly it computes, most of the results focusing on whether and when it converges and saying little about what it converges to. Another truly distributed algorithm for approximate inference is the popular Tree-Reweighted (TRW) version of BP \cite{wainwright}. In TRW, exact acyclic BP is performed on a sequence of spanning trees of the original cyclic network, where the potentials $\{\Upphi_i\}$ and $\{\Uppsi_{ij}\}$ are {\em reweighted} at each iteration in such a way as to ensure increasingly better inference. The method of Conditioning, discussed in the next section, can also be viewed as performing standard BP on a sequence of acyclic networks. 


\vspace{2mm}
\section{Conditioning for Undirected Networks}\label{sec:conditioning}

As mentioned in Section \ref{sec:introduction}, Conditioning \cite{pearl0} is an adaptation of BP for exact truly distributed message passing in cyclic networks. As with Clustering, it was initially studied in the context of directed networks. Let $L$ be a subset of nodes. The belief at node $i$ can be computed as
\begin{eqnarray}
  Z_i & = & \sum\limits_{\bfx_L} Z_i^{(\bfx_L)} , \label{eq:sum_conditioned_beliefs}
\end{eqnarray}
where $Z_i^{(\bfx_L)}$ is the belief at $i$ {\em conditioned on} the configuration $\bfx_L$. If $G$ is acyclic, then as in the previous section, one can use Belief Propagation to compute conditioned beliefs $Z_i^{(\bfx_L)}$ from conditioned messages $m_{j\rightarrow i}^{(\bfx_L)}$. If $G$ is cyclic, one can still use BP to compute beliefs by choosing $L$ to be a {\em loop cutset}, a set of nodes whose removal eliminates all cycles in the network. This is because conditioning on a node $l\in L$ effectively {\em splits} $l$ into multiple copies, each connected to a different subset of $l$'s neighbors $\partial l$, where the copies of $l$ are constrained to have the same value, the value upon which the original node $l$ is being conditioned. The process of splitting all nodes in $L$ can be interpreted as {\em opening up} the original cyclic network $G$ at the nodes in $L$. There was a great deal of research on finding loop cutsets in the context of directed networks \cite{becker}. The method of Conditioning can be viewed as performing BP on an opened up version of $G$ once for each configuration of the loop cutset $L$, and then combining the conditioned beliefs as in (\ref{eq:sum_conditioned_beliefs}). However, as with Clustering, actually creating a new network is possible only when the network and processing thereon resides on a centralized computer. Therefore, the opening of $G$ is just a schematic for visualizing node-specific messages that occur on the original cyclic network.

\begin{figure}[t]
	\begin{center}
		\hbox{
			\hspace{0.1in}
			\includegraphics[scale=.27]{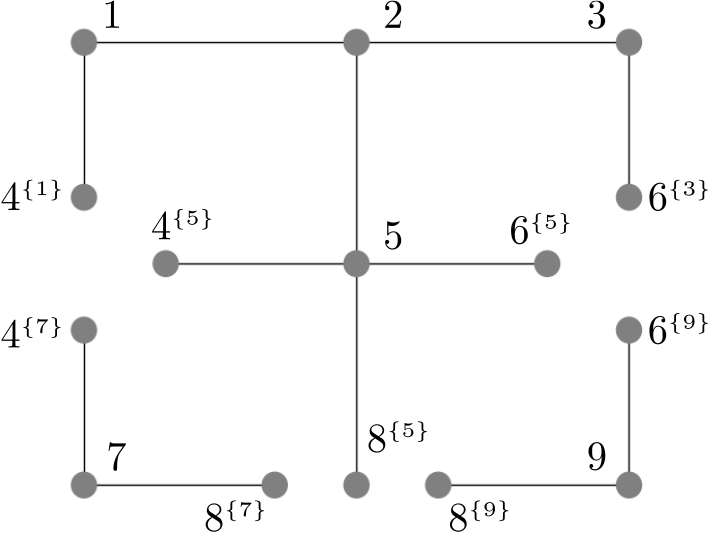}
			\hspace{0.3in}
			\includegraphics[scale=.27]{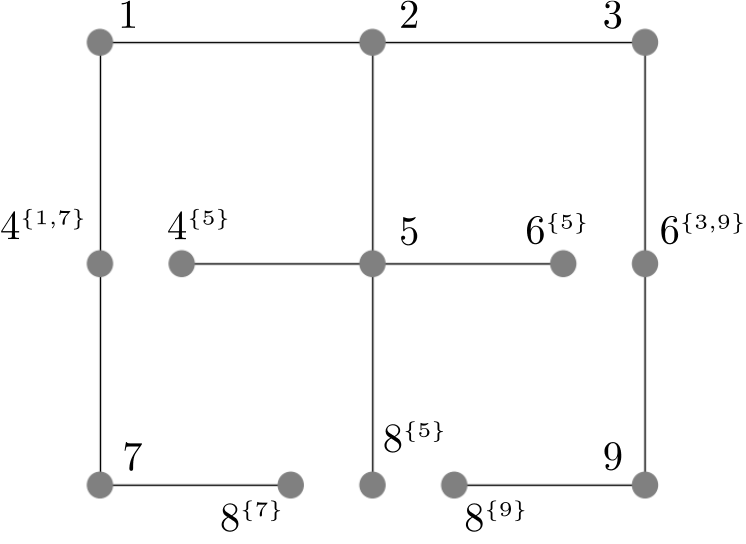}
		}
	\end{center}
	\vspace{-8mm}
	\hbox{\hspace{1.05in} (a) \hspace{2.2in} (b)}
	\caption{(a) Completely opened up $3\times 3$ grid network based on loop cutset $\{4,6,8\}$; (b) associated tree formed by re-identifying $4^{\{1\}}$ and $4^{\{7\}}$, and $6^{\{3\}}$ and $6^{\{9\}}$. Note that $\mathcal N_6 = \{3,9\}$ and $\mathcal L_6 = \{5\}$, while $\mathcal N_8 = \{ ~ \}$ and $\mathcal L_8 = \{7,5,9\}$.}
	\label{fig:assoc_tree}
\end{figure}

\vspace{3mm}
\subsection{\bf The Associated Tree and Node-Specific Message Passing}\label{sec:assoc_tree}

A procedure for finding an {\em opening} of $G$ based on a loop cutset $L$ is, for each $l\in L$, remove $l$ and all edges incident to $l$, then for each neighbor $j\in\partial l$, attach a {\em copy} $l^{\{j\}}$ of $l$ to $j$. As illustrated in Figure \ref{fig:assoc_tree} (a), the resulting network may be disconnected. It is relatively straightforward to show, however, that one can {\em reidentify} copies of a split loop cutset node to form a connected network that is still acyclic. This is illustrated in Figure \ref{fig:assoc_tree} (b). In principle one can use a connected or disconnected opening. However, a disconnected opening adds an additional layer of processing that cannot be performed in a truly distributed manner. For this reason one should consider a connected opening, which is referred to as the {\em associated tree} and denoted by $T$.

In standard BP all nodes form outgoing messages and compute beliefs in identical ways, using (\ref{eq:message_recursion}) and (\ref{eq:belief_formula}). In Conditioning, loop cutset nodes will form outgoing messages and compute beliefs from incoming messages differently depending on whether its neighbors are leaf or non-leaf nodes in $T$. For loop cutset node $l$, let $\mathcal{L}_l \subset \partial l$ denote the neighbors of $l$ for which the copy of $l$ attached to the neighbor is a leaf node in $T$. Loop cutset node $l$ will interact with $j$, in the original network, according to standard BP rules {\em as if} $l$ were a leaf connected only to $j$. On the other hand, let $\mathcal{N}_l = \partial l\setminus \mathcal{L}_l$ be the neighbors of $l$ that are connected to a non-leaf copy of $l$ in $T$. In this case, $l$ will interact with neighbors in $\mathcal N_l$ according to standard BP rules {\em as if} they were its only neighbors.

In order to correctly compute beliefs, a loop cutset node must use a consistent rule for dividing its self-potential among the messages it passes to neighbors in $\mathcal L_l$ and $\mathcal N_l$. In particular, it should use a different potential $\Upphi_{l^j} \defeq \Upphi_l^{\alpha_{j}}$ for each neighbor $j\in \mathcal L_l$, and another $\Upphi_{l^{\mathcal N_l}} \defeq \Upphi_l^{\alpha_{\mathcal N_l}}$ for all neighbors in $\mathcal N_l$, such that $\sum_{j\in\mathcal L_l} \alpha_j + \alpha_{\mathcal N_l} = 1$. To account for conditioning on a particular configuration $\bfx_L$ of the loop cutset, these self potentials will have to be modified, respectively, as $\Upphi^{(x_l)}_{l^{\mathcal N_l}} \defeq \Upphi_{l^{\mathcal N_l}} \delta^{(x_l)}$ and $\Upphi^{(x_l)}_{l^j} \defeq \Upphi_{l^j} \delta^{(x_l)}$, where $x_l$ is the value of node $l$ under loop cutset configuration $\bfx_L$.

Loop cutset node $l\in L$ passes to neighbor $j\in \mathcal L_l$ the message
\begin{eqnarray}
	m^{(\bfx_L)}_{l\rightarrow j} & = &  \Uppsi_{lj} \Upphi^{(x_l)}_{l^j} , \nonumber
\end{eqnarray}
while to a neighbor $j\in \mathcal{N}_l$, it passes the message
\begin{eqnarray}
	m^{(\bfx_L)}_{l\rightarrow j} & = &  \Uppsi_{lj} \Upphi^{(x_l)}_{l^{\mathcal N_l}} \prod\limits_{k\in\mathcal{N}_l\setminus j}m^{(\bfx_L)}_{k\rightarrow l} ~ . \nonumber
\end{eqnarray}

A loop cutset node $l\in L$ can compute conditioned beliefs from a neighbor $j\in \mathcal{L}_l$ as
\begin{eqnarray}
  Z^{(\bfx_L)}_{l} & = & \Upphi^{(x_l)}_{l^{j}} m^{(\bfx_L)}_{j\to l} ~ , \nonumber
\end{eqnarray}	
or from its non-leaf neighbors as
\begin{eqnarray}
  Z^{(\bfx_L)}_{l} & = & \Upphi^{(x_l)}_{l^{\mathcal N_l}} \prod\limits_{j\in\mathcal N_l} m^{(\bfx_L)}_{j\to l} ~ . \nonumber
\end{eqnarray}

For a non loop cutset node $i\not\in L$, the belief and outgoing messages conditioned on loop cutset configuration $\bfx_L$ are computed as
\begin{eqnarray}
Z^{(\bfx_L)}_{i} & = & \Upphi_j \prod\limits_{j\in \partial i} m^{(\bfx_L)}_{j\to i} \nonumber
\end{eqnarray}
and
\begin{eqnarray}
m^{(\bfx_L)}_{i\rightarrow j} & = & \Uppsi_{ij} \Upphi_i \prod\limits_{k\in\partial i\setminus j}m^{(\bfx_L)}_{k\rightarrow i} ~ . \nonumber
\end{eqnarray}

\vspace{3mm}
\subsection{\bf Parallel Implementation and Ordering}\label{sec:parallel}

Conditioning can be implemented both in serial and in parallel. However, the computational savings from Local Conditioning require parallel implementation. In this case, messages $\left[m^{(\bfx_L)}_{i\to j}\right]$ and beliefs $\left[Z_i^{(\bfx_L)}\right]$ corresponding to different loop cutset configurations are concatenated as columns in message and belief matrices $M^{(L)}_{i\to j}$ and $Z_i^{(L)}$, respectively. In the parallel implementation of Conditioning, there are $|\calX|^{|L|}$ columns in all message and belief matrices.

Parallel implementation requires that messages and beliefs are ordered so that corresponding columns of incoming messages themselves correspond to the same loop cutset configuration. We adopt the convention that column indices of message and belief matrices are ordered with respect to an ordering $(L)$ of $L$, given by their $|L|$-digit $|\calX|$-ary representations in which the first node of $(L)$ corresponds to the most significant digit, and so on. If the ordering $(L)$ is agreed upon in advance, the ordering of nodes in $L$ does not need to be communicated to neighboring nodes.

As with standard BP, for large networks messages in Conditioning will need to be scaled to avoid numerical underflow or overflow. If the columns of a message matrix $M^{(L)}_{i\to j}$ are scaled independently of one another, then an additional layer of non-distributed processing is required to compute beliefs for nodes in the network. Therefore, in order for beliefs to be computed in a truly distributed manner, all columns of a message matrix must be scaled with the same factor. We do not address this problem in the present paper.


\vspace{2mm}
\section{Local Conditioning for Undirected Networks}\label{sec:local}
\vspace{0mm}

{\em Local Conditioning} is an adaption of Conditioning that achieves exponential savings in complexity by conditioning message and belief matrices only on {\em local} subsets of the loop cutset nodes. Diez \cite{diez} introduced the main ideas of Local Conditioning for directed networks through examples, and Fay and Jaffray \cite{fay} subsequently proved that indeed exact beliefs can be computed with message matrices of reduced size. However, the demonstration in \cite{fay} was theoretical, showing that messages and beliefs {\em could be} computed only conditioning on local subsets of loop cutset nodes, and did not address the details of {\em how} nodes would actually compute beliefs and outgoing messages from incoming messages of different sizes and orderings. This section provides, in the context of undirected networks, both a more formal description of Local Conditioning than in \cite{diez}, and a more practical account than that given in \cite{fay}. In particular, it discusses the {\em summing out} of {\em upstream} loop cutset nodes when passing a message matrix over an edge, so that the message contains only columns corresponding to loop cutset nodes that are {\em relevant} for the edge; and the {\em expansion} of incoming message matrices to account for loop cutset nodes that are {\em downstream} of the incoming messages yet {\em relevant} for the receiving node. Furthermore, it discusses the {\em ordering} of message matrices with respect to the given relevant sets, and the {\em re}-ordering of incoming message matrices so that they are aligned prior to fusion.

Let $T$ be the associated tree dictating the conditioned message passing. Message passing in Local Conditioning still obeys the topology of $T$ and as such a loop cutset node $l$ computes outgoing messages to neighbors, and beliefs from incoming messages, depending on whether $l$ is connected to its neighbors as a leaf or non-leaf in $T$. To simplify the ensuing discussion, message and belief computations will be presented agnostically, without reference to whether a node is in the loop cutset or not. However, it should be clear from the discussion in Section \ref{sec:assoc_tree} that if one wishes to specifically consider a loop cutset node, either the set $\mathcal N_l$, or $\{j\}$ for some $j\in \mathcal L_l$, can be substituted for $\partial i$. Moreover, care will of course need to be applied to self-potentials of loop cutset nodes, both in dividing them up between $\mathcal L_l$ and $\mathcal N_l$, and in modifying them to ensure consistency with the respective loop cutset configurations.

\vspace{3mm}
\subsection{\bf Relevant Nodes and Reduced Complexity}

Recall from Section \ref{sec:bp} that a node $k$ is {\em upstream} of the message $m_{j\to i}$ if $k \in T_{j\setminus i}$. Likewise, $k$ is said to be {\em downstream} of $m_{j\to i}$ if $k\in T_{i\setminus j}$. Note that if $k$ is upstream of $m_{j\to i}$, then $k$ is downstream of $m_{i\to j}$. Let $L_{j\setminus i} \subset L$ denote the subset of loop cutset nodes {\em all of whose copies} are contained in $T_{j\setminus i}$, or in other words, the set of loop cutset nodes upstream of $m_{j\to i}$ and downstream of $m_{i\to j}$. Note that $L_{j\setminus i}$ and $L_{i\setminus j}$ are disjoint. For an edge $\{i,j\}$, the {\em relevant set}
\begin{eqnarray}
    R_{ij} ~~ \defeq ~~ L\setminus \left(L_{i\setminus j} \cup L_{j\setminus i}\right) \label{eq:edge_rel}
\end{eqnarray}
is the set of loop cutset nodes at least one copy of which is upstream and at least one copy of which is downstream of the messages $m_{j\to i}$ and $m_{i\to j}$ passed over the edge. Clearly, $R_{ij}$ and $R_{ji}$ are the same. For node $i$, the relevant set is
\begin{eqnarray}
    R_i ~ = \bigcup\limits_{j\in\partial i}R_{ij} ~ . \label{eq:node_rel}
\end{eqnarray}
A message matrix $M^{(R_{ji})}_{j\to i}$ passed over edge $\{i,j\}$ will only be conditioned on the relevant set $R_{ij}$, and as such will have $|\calX|^{|R_{ij}|}$ columns, one for each configuration $\bfx_{R_{ij}}$ on the relevant set for the edge. When node $i$ receives message matrices $M_{j\to i}^{(R_{ji})}$ from its neighbors, each will be conditioned on a potentially different subset of the relevant set $R_i$ for node $i$. Node $i$ will then {\em expand} each of the incoming $M_{j\to i}^{(R_{ji})}$ to a message matrix $M_{j\to i}^{(R_i)}$ conditioned on node $i$'s relevant set. After expanding each of the incoming message matrices, node $i$ will then {\em reorder} the expanded message matrices so that a given column of the incoming messages correspond to the same configuration $\bfx_{R_i}$. The belief matrix $Z_i^{(R_i)}$ is then computed by multiplying element-wise the reordered incoming messages and the self-potential for $i$ as
\begin{eqnarray}
    Z_{i}^{(\bfx_{R_i})} & = & \Upphi_i \prod\limits_{j\in \partial i} m_{j\to i}^{(\bfx_{R_{ji}}, \bfx_{R_i\setminus R_{ji}})} ~ , \nonumber
\end{eqnarray}
and then computing unconditioned beliefs as
\begin{eqnarray}
    Z_i & = & \sum\limits_{x_{R_i}}Z_i^{(\bfx_{R_i})} ~ . \nonumber
\end{eqnarray}
Node $i$ will also form the outgoing message column
\begin{eqnarray}
    ~~~ m_{i\to j}^{(\bfx_{R_{ij}})}  & = &  \Uppsi_{ji} \Upphi_i \sum\limits_{\bfx_{R_i \setminus R_{ij}}}\prod\limits_{k\in\partial i\setminus j} m_{k\to i}^{(\bfx_{R_{ki}}, \bfx_{R_i \setminus R_{ki}})} ~ , \nonumber 
\end{eqnarray}
to neighbor $j$ by multiplying element-wise expanded and reordered message matrices from its other neighbors $k\in\partial i\setminus j$, and then {\em summing out} those loop cutset nodes in $R_i$ but not in $R_{ij}$.


\vspace{3mm}
\subsection{The Details}

Details of {\em summing out}, {\em expanding}, and {\em reordering} for Local Conditioning are discussed. For a given node $i$ in $T$, let $j_1,j_2,\ldots, j_{n_i}$ indicate the neighbors of $i$. $L$ can be partitioned as $(R_i,L_{j_1\setminus i},\ldots,L_{j_{n_i}\setminus i})$ In Conditioning, the message $M_{j_m\to i}^{(L)}$ from $j_m$ to $i$ is a matrix with $|\calX|^{|L|}$ columns, one for each configuration of the loop cutset $L$. By definition, $L$ can be partitioned as $(L_{i\setminus j_m},R_{i,j_m},L_{j_m\setminus i})$. 

In Local Conditioning, the message matrix $M_{j_m\to i}^{(L_{i\setminus j_m},R_{i,j_m},L_{j_m\setminus i})}$ need not be conditioned on $\bfx_{L_{i\setminus j_m}}$. For any two configurations $(\bfx_{L_{i\setminus j_m}}, \bfx_{R_{j_m,i}}, \bfx_{L_{j_m\setminus i}})$ and $(\bfx'_{L_{i\setminus j_m}}, \bfx_{R_{j_m,i}}, \bfx_{L_{j_m\setminus i}})$ that differ only on $L_{i\setminus j_m}$, the corresponding message columns $m_{j_m\to i}^{(\bfx_{L_{i\setminus j_m}}, \bfx_{R_{j_m,i}}, \bfx_{L_{j_m\setminus i}})}$ and $m_{j_m\to i}^{(\bfx'_{L_{i\setminus j_m}}, \bfx_{R_{j_m,i}}, \bfx_{L_{j_m\setminus i}})}$ are identical. This is because all potentials involving nodes in $L_{i\setminus j_m}$ are downstream of the message from $j_m$ to $i$. We can eliminate this redundancy, and node $j_m$ can pass to node $i$ a message matrix $M_{j_m\to i}^{(R_{j_m,i},R_{L_{j_m\setminus i}})}$ with only $|\calX|^{|R_{j_mi}| + |R_{L_{j_m\setminus i}}|}$ columns.

Considering the message matrix $M^{(R_{j_m,i},R_{L_{j_m\setminus i}})}$, node $j_m$ can sum the columns 
\begin{eqnarray}
    \{m_{j_m\to i}^{(\bfx_{R_{i,j_m}},\bfx'_{L_{j_m\setminus i}})} : \bfx'_{L_{j_m\setminus i}} \in \calX_{L_{j_m\setminus i}}\} \nonumber
\end{eqnarray}
of message matrix $M_{j_m\to i}^{(R_{i,j_m}, L_{j_m\setminus i})}$ that agree on $R_{i,j_m}$ and differ on $L_{j_m\setminus i}$. That is, since all potentials involving nodes in $L_{j_m\setminus i}$ are upstream of the message from $j_m$ to $i$, it is not necessary to retain separate columns for different configurations on the these loop cutset nodes. This results in a message matrix $M_{j_m\to i}^{(R_{i,j_m})}$ with $|\calX|^{|R_{i,j_m}|}$ columns, a substantial reduction from the original number of $|\calX|^{|L|}$.

As in Section \ref{sec:parallel}, columns of the message $M_{j_m\to i}^{(R_{j_m,i})}$ are ordered with respect to an ordering $(R_{j_m,i})$ of the relevant set $R_{j_m,i}$ for edge $\{i,j_m\}$. Each incoming $M_{j_m\to i}^{(R_{j_m,i})}$ is expanded to a message $M_{j_m\to i}^{(R_{i})}$ by iteratively replicating $M_{j_m\to i}^{(R_{j_m,i})}$ for each node in $R_i \setminus R_{i,j_m}$. Let $(R_i)_{j_m}$ denote the ordering of $R_i$ with respect to which the columns of expanded $M^{(R_i)}_{j_m\to i}$ are ordered, and $(R_i)$ the common ordering of $R_i$ with which all incoming message matrices need to be aligned. For each neighbor $j_m\in\partial i$, node $i$ computes $P_{(R_{i})_{j_m}}$, the permutation matrix that converts $(R_i)_{j_m}$ to $(R_i)$. Using $P_{(R_{i})_{j_m}}$, indices of the expanded $M^{(R_i)_{j_m}}_{j_m\to i}$ are mapped to the reordered $M^{(R_i)}_{j_m\to i}$ in the following way. A column $m^{(\bfx_{R_i})_{j_m}}_{j_m\to i}$ of expanded $M^{(R_i)_{j_m}}_{j_m\to i}$ has an index $c = (c_{1} \cdots c_{|R_i|})$, where $c_k\in\{0,\ldots,|\calX|\}$. Applying $P_{(R_{i})_{j_m}}$ to $c$ gives the index within reordered $M^{(R_i)}_{j_m\to i}$ for the conditioned message $m^{(\bfx_{R_i})}_{j_m\to i}$.

We stated above that node $j_m$ sums columns of message matrix $M_{j_m\to i}^{(R_{j_m,i},L_{j_m\setminus i})}$ corresponding to different configurations on $L_{j_m\setminus i}$, resulting in a message matrix $M_{j_m\to i}^{(R_{j_m,i})}$ of reduced size. In reality, the summing out of loop cutset nodes in $L_{j_m\setminus i}$ will have been performed recursively, neighbors $k\in\partial j_m\setminus i$ summing out some of the nodes in $L_{j_m\setminus i}$, and so on back to the leaves of $T_{j_m\setminus i}$.

For a given neighbor $k\in\partial j_m\setminus i$, the set $R_{j_m,i} \cup L_{j_m\setminus i}$ can equivalently be partitioned as $(L^{j_m\setminus i}_{j_m\setminus k} , R_{k,j_m}, L_{k\setminus j_m})$, where $L^{j_m\setminus i}_{j_m\setminus k} \defeq L_{j_m\setminus i} \cap L_{j_m\setminus k}$ is the set of loop cutset nodes all of whose copies are upstream of the message from $j_m$ to $i$ and downstream of the message from $k$ to $j_m$. For example, consider {\em another} neighbor $k'\in\partial j_m\setminus \{i,k\}$. The set $L_{k'\setminus j_m}$ consists of loop cutset nodes all of whose copies are in both $T_{j_m\setminus i}$ and $T_{j_m\setminus k}$. Using similar arguments as before, the message $M_{k\to j_m}^{(\bfx_{L^{j_m\setminus i}_{j_m\setminus k}} ~ , ~ \bfx_{R_{k,j_m}} ~ , ~ \bfx_{L_{k\setminus j_m}})}$ need not be conditioned on $L^{j_m\setminus i}_{j_m\setminus k}$, and the loop cutset nodes in $L_{k\setminus j_m}$ can be summed out {\em before} node $k$ sends its message to $j_m$. As a result, node $j_m$ receives from each $k\in\partial j_m\setminus i$ a message $M_{k\to j_m}^{(R_{k,j_m})}$. The remaining nodes in $L_{j_m\setminus i}$ to be summed out before node $j_m$ passes its message to $i$ are those loop cutset nodes that are relevant for at least one $k\in\partial j_m\setminus i$ and upstream of the message from $j_m$ to $i$, which we denote by $L^{j_m\setminus i} ~ \defeq \left(\cup_{k\in\partial j_m\setminus i} R_{k,j_m}\right) \setminus R_{i,j_m}$. That is, the incoming message matrices $M_{k\to j_m}^{(R_{k,j_m})}$, $k\in\partial j_m\setminus i$, are each expanded to $\cup_{k\in\partial j_m\setminus i} R_{k,j_m}$, reordered so that corresponding columns align, then multiplied element-wise along with the self-potential $\Upphi_{j_m}$ for $j_m$. Finally, conditioning on $L^{j_m\setminus i}$ is removed by summing out, resulting in the outgoing message matrix $M_{j_m\to i}^{(R_{i,j_m})}$.

We now show the algebraic details of computing beliefs and messages.

\vspace{3mm}
\subsubsection{\bf Computing Beliefs}

The belief $Z_i$ of node $i$ can be expressed as
\begin{eqnarray}
  Z_i & = & \sum\limits_{\bfx_L} Z_i^{(\bfx_L)} \nonumber \\
      & = & \sum\limits_{\bfx_L} \Upphi_i \prod\limits_{j\in \partial i} m_{j\rightarrow i}^{(\bfx_L)} \nonumber \\
      & = & \sum\limits_{\bfx_{R_i}} \sum\limits_{\bfx_{L_{j_1\setminus i}}} \cdots \sum\limits_{\bfx_{L_{j_{n_i}\setminus i}}} \Upphi_i \prod\limits_{m=1}^{n_i} m_{j_m\rightarrow i}^{(\bfx_L)} \nonumber \\
      & = & \sum\limits_{\bfx_{R_i}} \Upphi_i \sum\limits_{\bfx_{L_{j_1\setminus i}}} \cdots \sum\limits_{\bfx_{L_{j_{n_i}\setminus i}}}  m_{j_1\rightarrow i}^{(\bfx_L)} \cdots m_{j_{n_i}\rightarrow i}^{(\bfx_L)} \nonumber \\
      & = & \sum\limits_{\bfx_{R_i}} \Upphi_i \sum\limits_{\bfx_{L_{j_1\setminus i}}} \cdots \sum\limits_{\bfx_{L_{j_{n_i}\setminus i}}}  m_{j_1\rightarrow i}^{(\bfx_{L_{i \setminus j_1}},\bfx_{R_{i,j_1}}, \bfx_{L_{j_1 \setminus i}})} \cdots m_{j_{n_i}\rightarrow i}^{(\bfx_{L_{i \setminus j_{n_i}}}, \bfx_{R_{i,j_{n_i}}}, \bfx_{L_{j_{n_i} \setminus i}})} ~ . \nonumber
\end{eqnarray}
Recall from preceding discussion, that the message $m_{j_1\to i}^{(\bfx_L)}$ is not a function of $\bfx_{L_{j_2\setminus i}}, \ldots, \bfx_{L_{j_{n_i}\setminus i}}$. Therefore, each message $m_{j_m\to i}^{(\bfx_{L_{i \setminus j_{m}}}, \bfx_{R_{i,j_{m}}}, \bfx_{L_{j_{m} \setminus i}})}$ becomes $m_{j_m\to i}^{(\bfx_{R_{i,j_{m}}}, \bfx_{L_{j_{m} \setminus i}})}$, the summations $\sum_{L_{j_1\setminus i}} \cdots \sum_{L_{j_{n_i}\setminus i}}$ distribute over the multiplications $m_{j_1\to i}^{(\bfx_{R_{i,j_{1}}}, \bfx_{L_{{j_1} \setminus i}})} \cdots m_{j_{n_i}\to i}^{(\bfx_{R_{i,j_{n_i}}}, \bfx_{L_{j_{n_i} \setminus i}})}$, and we continue as
\begin{eqnarray}
   Z_i   & = & \sum\limits_{\bfx_{R_i}} \Upphi_i \sum\limits_{\bfx_{L_{j_1\setminus i}}} m_{j_1\to i}^{(\bfx_{R_{i,j_1}}, \bfx_{L_{j_1 \setminus i}})} \cdots \sum\limits_{\bfx_{L_{j_{n_i}\setminus i}}} m_{j_{n_i}\to i}^{(\bfx_{R_{i,j_{n_i}}}, \bfx_{L_{j_{n_i} \setminus i}})} \nonumber \\
      & = & \sum\limits_{\bfx_{R_i}} \Upphi_i \prod\limits_{m=1}^{n_i} \sum\limits_{\bfx_{L_{j_m\setminus i}}} m_{j_m\to i}^{(\bfx_{R_{j_m,i}}, \bfx_{L_{j_m \setminus i}})} \label{eq:pre_sumout} \\
      & = & \sum\limits_{\bfx_{R_i}} \Upphi_i \prod\limits_{m=1}^{n_i} m_{j_m\to i}^{(\bfx_{R_{j_m,i}})} ~ . \nonumber
\end{eqnarray}
{\em Summing out} loop cutset nodes in the $L_{j_m\setminus i}$ results in incoming message matrices $M_{j_m\to i}^{(R_{j_m,i})}$ with $|\calX|^{|R_{j_m,i}|}$ columns, respectively.


After receiving incoming messages $M_{j\to i}^{({R_{ji}})}$ from its neighbors $j\in\partial i$, node $i$ {\em expands} each of them to matrices $M_{j\to i}^{({R_{i}})_j}$ conditioned on its relevant set $R_i$ by duplicating the columns of $M_{j\to i}^{({R_{ji}})}$ repeatedly for each node in $R_i \setminus R_{ij}$. The expanded message matrices are then {\em reordered} so that they can be fused through element-wise multiplication. Node $i$ then computes its belief as
\begin{eqnarray}
      Z_i & = & \sum\limits_{\bfx_{R_i}} \Upphi_i \prod\limits_{m=1}^{n_i} m_{j_m\to i}^{(\bfx_{R_i})} ~ , \label{eq:unconditioned_belief_local}
\end{eqnarray}
which corresponds to the operation of node $i$ summing the $|\calX|^{|R_i|}$ columns of its belief matrix $Z_i^{(R_i)}$, the columns of which correspond to different configurations $\bfx_{R_i}$, formed by multiplying element-wise its self-potential $\Upphi_i$ and the incoming message matrices $M_{j\to i}^{(R_i)}$ from its neighbors.


\vspace{3mm}
\subsubsection{\bf Computing Outgoing Messages}


In the transition from (\ref{eq:pre_sumout}) to (\ref{eq:unconditioned_belief_local}) above, we see that node $j$ passes to node $i$ the message
\begin{eqnarray}
    m_{j\to i}^{(\bfx_{R_{ji}})} & = & \sum\limits_{\bfx_{L_{j\setminus i}}} m_{j\to i}^{(\bfx_{R_{ji}} ~ , ~ \bfx_{L_{j \setminus i}})} \nonumber \\
			  & = & \sum\limits_{\bfx_{L_{j\setminus i}}} \Uppsi_{j,i} \Upphi_{j} \prod\limits_{k\in\partial j\setminus i} m_{k\to j}^{(\bfx_{R_{ji}} ~ , ~ \bfx_{L_{j \setminus i}})} \nonumber \\
			  & = & \Uppsi_{j,i} \Upphi_{j}  \sum\limits_{\bfx_{L^{j\setminus i}}} \sum\limits_{\bfx_{L_{k_1\setminus j}}}  \cdots \sum\limits_{\bfx_{L_{k_{n_j}\setminus j}}} m_{k_1\to j}^{(\bfx_{R_{ji}} ~ , ~ \bfx_{L_{j \setminus i}})} \cdots m_{k_{n_j}\to j}^{(\bfx_{R_{ji}} ~ , ~ \bfx_{L_{j \setminus i}})} \nonumber \\
			  & = & \Uppsi_{j,i} \Upphi_{j}  \sum\limits_{\bfx_{L^{j\setminus i}}} \prod\limits_{k_m\in\partial j\setminus i} \sum\limits_{\bfx_{L_{k_{m}\setminus j}}} m_{k_m\to j}^{(\bfx_{R_{ji}} ~ , ~ \bfx_{L_{j \setminus i}})}  \nonumber \\
			  & = & \Uppsi_{j,i} \Upphi_{j}  \sum\limits_{\bfx_{L^{j\setminus i}}} \prod\limits_{k_m\in\partial j\setminus i} \sum\limits_{\bfx_{L_{k_{m}\setminus j}}} m_{k_m\to j}^{(\bfx_{L_{j\setminus k_m}^{j\setminus i}} , ~ \bfx_{R_{k_m,j}} ~ , ~ \bfx_{L_{k_m \setminus j}})}  \nonumber
\end{eqnarray}
where we have used the fact, mentioned above, that for a neighbor $k\in\partial j\setminus i$, $R_{ji} \cup L_{j\setminus i}$ can be partitioned as $(L^{j\setminus i}_{j\setminus k},R_{kj},L_{k\setminus j})$. As before, note that the set $L_{j\setminus k}^{j\setminus i}$ is downstream of the message from $k$ to $j$ and as such for any two configurations $\bfx_{L^{j\setminus i}_{j\setminus k}}$ and $\bfx'_{L^{j\setminus i}_{j\setminus k}}$, the messages $m_{k\to j}^{(\bfx_{L_{j\setminus k}^{j\setminus i}} , ~ \bfx_{R_{k,j}} ~ , ~ \bfx_{L_{k \setminus j}})}$ and $m_{k\to j}^{(\bfx'_{L_{j\setminus k}^{j\setminus i}} , ~ \bfx_{R_{k,j}} ~ , ~ \bfx_{L_{k \setminus j}})}$ are identical. Therefore the message from $k$ to $j$ does not need to be conditioned on $L_{j\setminus k}^{j\setminus i}$, leaving us with
\begin{eqnarray}
	m_{j\to i}^{(\bfx_{R_{ji}})} & = & \Uppsi_{j,i} \Upphi_{j}  \sum\limits_{\bfx_{L^{j\setminus i}}} \prod\limits_{k_m\in\partial j\setminus i} \sum\limits_{\bfx_{L_{k_{m}\setminus j}}} m_{k_m\to j}^{(\bfx_{R_{k_m,j}} ~ , ~ \bfx_{L_{k_m \setminus j}})}  \nonumber \\
					  & = & \Uppsi_{j,i} \Upphi_{j}  \sum\limits_{\bfx_{L^{j\setminus i}}} \prod\limits_{k_m\in\partial j\setminus i} m_{k_m\to j}^{(\bfx_{R_{k_m,j}})} . \nonumber 
\end{eqnarray}

Node $j$ {\em expands} each of the incoming message matrices $M^{(R_{k_m,j})}_{k_m\to j}$ to message matrix $M^{(\cup_{k\in\partial j\setminus i} R_{kj})_{k_m}}_{k_m\to j}$, {\em reorders} the expanded message matrices to facilitate element-wise multiplication, then sums out the loop cutset nodes in $L^{j\setminus i} = R_j \setminus R_{ij}$. This yields the outgoing message matrix $M^{(R_{ji})}_{j\to i}$ and establishes recursive computation of Local Conditioning messages.

We summarize belief and message computation in the following theorem.

\vspace{6mm}
\begin{theorem}[Local Conditioning Sum-Product BP]
	
	For a non loop cutset node $j\not\in L$, local conditioned beliefs and messages are computed as
	\begin{eqnarray}
	Z_{j}^{(\bfx_{R_j})} & = & \Upphi_j \prod\limits_{k\in \partial j} m_{k\to j}^{(\bfx_{R_{kj}})} \nonumber
	\end{eqnarray}
	and
	\begin{eqnarray}
	m_{j\to i}^{(\bfx_{R_{ji}})}  & = &  \Uppsi_{ji} \Upphi_j \sum\limits_{\bfx_{L^{j\setminus i}}} \prod\limits_{k\in\partial j\setminus i} m_{k\to j}^{(\bfx_{R_{jk}})} , \nonumber 
	\end{eqnarray}
	where $L^{j\setminus i} ~ \defeq \left(\cup_{k\in\partial j\setminus i} R_{kj}\right) \setminus R_{ij}$ is the set of loop cutset nodes all of whose copies are upstream of the message from $j$ to $i$ but are 
	not in any of the $L_{k\setminus j}$ for $k\in \partial j\setminus i$. Unconditioned beliefs are computed as
	\begin{eqnarray}
	Z_j & = & \sum\limits_{\bfx_{R_j}}Z_j^{(\bfx_{R_j})}. \nonumber
	\end{eqnarray}
	
	\vspace{2mm}
	
	For a loop cutset node $l\in L$, conditioned beliefs are computed as
	\begin{eqnarray}
	Z^{(\bfx_{R_l})}_{l} & = & \Upphi_{l^j}^{(x_{l})} m^{(\bfx_{R_{jl}})}_{j\to l}, \nonumber
	\end{eqnarray}	
	for some $j\in\mathcal{L}_l$, or as
	\begin{eqnarray}
	Z^{(\bfx_{R_l})}_{l} & = & \Upphi_{l^{\mathcal N_l}}^{(x_{l})} \prod\limits_{j\in\mathcal N_l} m^{(\bfx_{R_{jl}})}_{j\to l} . \nonumber
	\end{eqnarray}	
	The message to a neighbor $i\in\mathcal{N}_l$ is computed as
	\begin{eqnarray}
	m^{(\bfx_{R_{li}})}_{l\rightarrow i} & = &  \Uppsi_{li} \Upphi_{l^{\mathcal N_l}}^{(x_{l})} \sum\limits_{\bfx_{L^{l\setminus i}}} \prod\limits_{k\in\mathcal{N}_l\setminus i} m^{(\bfx_{R_{kl}})}_{k\rightarrow l}, \nonumber
	\end{eqnarray}
	where $L^{l\setminus i} ~ \defeq \left(\cup_{k\in\partial l\setminus i} R_{kl}\right) \setminus R_{il}$ is the set of loop cutset nodes all of whose copies are upstream of the message from $l$ to $i$ but are not in any of the $L_{k\setminus l}$ for $k\in \partial l\setminus i$.  Unconditioned beliefs are computed as
	\begin{eqnarray}
		Z_l & = & \sum\limits_{\bfx_{R_{l}}}Z_l^{(\bfx_{R_{l}})}. \nonumber
	\end{eqnarray}
\end{theorem}

\vspace{3mm}
\subsection{Complexity of Local Conditioning}

The complexity of Local Conditioning depends upon the sizes of the relevant sets for nodes and edges with respect to the particular associated tree used for the node-specific message passing of LC. For a given loop cutset $L$, different associated trees will result in different computational complexities. We have not addressed the question of finding an optimal associated tree for a given loop cutset, nor the further question of finding an optimal loop cutset. Considerable work has been done in optimizing loop cutsets in the context of directed networks \cite{becker}, and our hope is that some of this can be leveraged for the undirected case.

For a given associated tree, we have shown that the message passed over edge $\{i,j\}$ is a matrix with $|\calX|^{|R_{ij}|}$ columns. Each column has $|\calX|$ elements. Moreover, the belief at node $i$ is a matrix with $|\calX|^{|R_i|}$ columns, again each column with $|\calX|$ elements. Since $R_{ij} \subset R_i$ for all $j\in\partial i$, the complexity is dominated by a term that is exponential in $\max_{i\in V} |R_i|$. In our preliminary analysis so far, we have found that for an $M\times N$ grid network, there is a loop cutset and an associated tree for that loop cutset with $\max_i |R_i| = M+1$. In other words, the complexity of Local Conditioning on an $M\times N$ grid network is the same as that of Clustering.

\vspace{3mm}
\subsection{An Example: Expanding, Reordering, and Summing Out}

\begin{figure}[t]
	\begin{center}
		\hbox{
			\hspace{0.1in}
			\includegraphics[scale=.27]{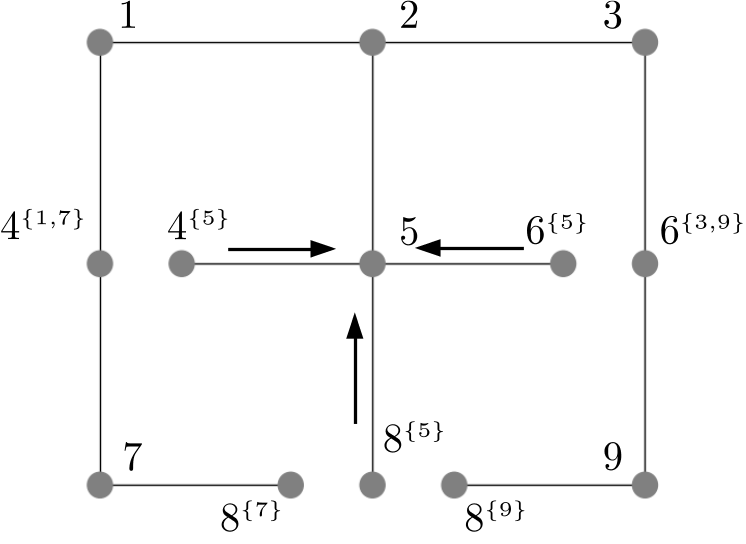}
			\hspace{0.3in}
			\includegraphics[scale=.27]{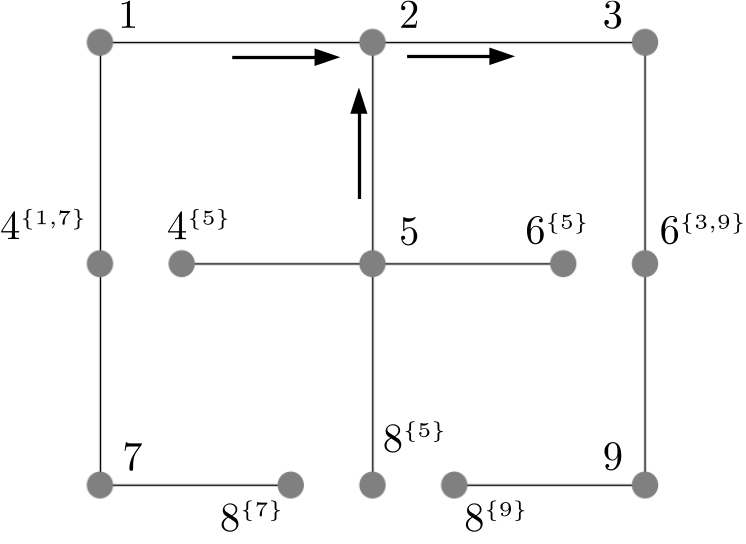}
		}
	\end{center}
	\vspace{-8mm}
	\hbox{\hspace{1.05in} (a) \hspace{2.2in} (b)}
	\caption{(a) Messages $M^{(4)}_{4\to 5}$, $M^{(8)}_{8\to 5}$, and $M^{(6)}_{6\to 5}$ incoming to node $5$ from neighbors $4$, $8$, and $6$, respectively; (b) outgoing message $M^{(6,8)}_{2\to 3}$ computed from $M^{(4,8)}_{1\to 2}$ and $M^{(4,6,8)}_{5\to 2}$.}
	\label{fig:assoc_tree_lc}
\end{figure}

Figure \ref{fig:assoc_tree_lc} illustrates Local Conditioning messages passed with respect to the associated tree from Figure \ref{fig:assoc_tree}. In (a), node 5 receives incoming message matrices 
\begin{align*}
M_{4\to 5}^{(4)} & = \left[\begin{array}{cc}
			m^{(-1)}_{4\to 5} , & m^{(1)}_{4\to 5} \\
		    \end{array}\right] , & \!\!  M_{6\to 5}^{(6)} & = \left[\begin{array}{cc}
									  m^{(-1)}_{6\to 5} , & m^{(1)}_{6\to 5} \\
								      \end{array}\right] , & \!\!  M_{8\to 5}^{(8)} & = \left[\begin{array}{cc}
													    m^{(-1)}_{8\to 5} , & m^{(1)}_{8\to 5} \\
													\end{array}\right]
\end{align*}                                                                   
from neighbors $4$, $6$, and $8$, respectively. Each of these is expanded to a message matrix conditioned on $\{4,6,8\}$ in the following way:
\begin{eqnarray}
 M_{4\to 5}^{(6,8,4)} & = \left[\begin{array}{cccccccc}
			m^{(-1)}_{4\to 5} ~ , & m^{(1)}_{4\to 5} ~ , & m^{(-1)}_{4\to 5} ~ , & m^{(1)}_{4\to 5} ~ , & m^{(-1)}_{4\to 5} ~ , & m^{(1)}_{4\to 5} ~ , & m^{(-1)}_{4\to 5} ~ , & m^{(1)}_{4\to 5} \\
		    \end{array}\right] , \nonumber
\end{eqnarray}
\begin{eqnarray}
 M_{6\to 5}^{(4,8,6)} & = \left[\begin{array}{cccccccc}
			m^{(-1)}_{6\to 5} ~ , & m^{(1)}_{6\to 5} ~ , & m^{(-1)}_{6\to 5} ~ , & m^{(1)}_{6\to 5} ~ , & m^{(-1)}_{6\to 5} ~ , & m^{(1)}_{6\to 5} ~ , & m^{(-1)}_{6\to 5} ~ , & m^{(1)}_{6\to 5} \\
		    \end{array}\right] , \nonumber
\end{eqnarray}
and
\begin{eqnarray}
 M_{8\to 5}^{(4,6,8)} & = \left[\begin{array}{cccccccc}
			m^{(-1)}_{8\to 5} ~ , & m^{(1)}_{8\to 5} ~ , & m^{(-1)}_{8\to 5} ~ , & m^{(1)}_{8\to 5} ~ , & m^{(-1)}_{8\to 5} ~ , & m^{(1)}_{8\to 5} ~ , & m^{(-1)}_{8\to 5} ~ , & m^{(1)}_{8\to 5} \\
		    \end{array}\right] . \nonumber
\end{eqnarray}
These expanded message matrices will be ordered with respect to the ordering $(4,6,8)$ of $R_5$. The expanded messages $M^{(6,8,4)}_{4\to 5}$ and $M^{(4,8,6)}_{6\to 5}$ from nodes $4$ and $6$ are reordered using the permutation matrices
\begin{align*}
  P_{(6,8,4)} & = \left[\begin{array}{ccc}
                           0 & 0 & 1 \\
                           1 & 0 & 0 \\
                           0 & 1 & 0 \\
                          \end{array}\right] ~~~~~~~~~~~~~~ , & P_{(4,8,6)} & = \left[\begin{array}{ccc}
									  1 & 0 & 0 \\
									  0 & 0 & 1 \\
									  0 & 1 & 0 \\
									  \end{array}\right] ~ ,
\end{align*}
that respectively permute $(6,8,4)$ and $(4,8,6)$ into $(4,6,8)$. That is, columns of $M^{(6,8,4)}_{4\to 5}$ are indexed as $000, 001, 010, 011, 100, 101, 110$ and $111$. Multiplying each of these indices by $P_{(6,8,4)}$ shows that the columns of $M^{(6,8,4)}_{4\to 5}$ are respectively reindexed as in
\begin{eqnarray}
  M_{4\to 5}^{(4,6,8)} & = & \left[\begin{array}{cccccccc}
				    m^{(-1)}_{4\to 5} ~ , & m^{(1)}_{4\to 5} ~ , & m^{(-1)}_{4\to 5} ~ , & m^{(1)}_{4\to 5} ~ , & m^{(-1)}_{4\to 5} ~ , & m^{(1)}_{4\to 5} ~ , & m^{(-1)}_{4\to 5} ~ , & m^{(1)}_{4\to 5} \\
				\end{array}\right] ~ . \nonumber
\end{eqnarray}
Likewise for $M^{(4,8,6)}_{6\to 5}$ and $P_{(4,8,6)}$.

In (b), node 2 forms preliminary outgoing message 
\begin{eqnarray}
 M_{2\to 3}^{(4,6,8)} & = \left[\begin{array}{cccccccc}
			\bar m^{(0)}_{4\to 5} ~ , & \bar m^{(1)}_{4\to 5} ~ , & \bar m^{(2)}_{4\to 5} ~ , & \bar m^{(3)}_{4\to 5} ~ , & \bar m^{(4)}_{4\to 5} ~ , & \bar m^{(5)}_{4\to 5} ~ , & \bar m^{(6)}_{4\to 5} ~ , & \bar m^{(7)}_{4\to 5} \\
		    \end{array}\right] ~ , \nonumber
\end{eqnarray}
where the configuration on $\{4,6,8\}$ is indicated by enumerating the combinations of $\{-1,1\}^3$, the binary representation of a configuration index having $4$ has the most significant digit. Since we want to sum out node $4$, node 2 will form the outgoing message matrix $M_{2\to 3}^{(6,8)}$ in the following way:
\begin{eqnarray}
  M_{2\to 3}^{(6,8)} & = & M_{2\to 3}^{(4,6,8)}\left[\begin{array}{cccc}
                                                      1 & 0 & 0 & 0 \\
                                                      0 & 1 & 0 & 0 \\
                                                      0 & 0 & 1 & 0 \\
                                                      0 & 0 & 0 & 1 \\
                                                      1 & 0 & 0 & 0 \\
                                                      0 & 1 & 0 & 0 \\
                                                      0 & 0 & 1 & 0 \\
                                                      0 & 0 & 0 & 1 \\
                                                     \end{array}\right] ~ . \nonumber
\end{eqnarray}

\vspace{1mm}
\section{Concluding Remarks and Future Directions}\label{sec:conclusion}

We have addressed issues related to the practical implementation of Local Conditioning. As more and more systems are fielded comprised of distributed entities, such as sensor networks or roadways of autonomous vehicles, an understanding of Local Conditioning will become increasingly important. Further work in this area must address suboptimal implementations for networks of prohibitive topology, for example by prematurely summing out relevant loop cutset nodes.

\section*{Acknowledgements}
The author would like to thank David Neuhoff for comments on an earlier draft.

%
%

\end{document}